**Large epitaxial bi-axial strain induces a Mott-like phase transition in $VO_2$**

*Salinporn Kittiwantanakul, Stuart A. Wolf, Jiwei Lu**

Salinporn Kittiwatanakul
Department of Physics, University of Virginia, Charlottesville, VA, U.S., 22904
Prof. S. A. Wolf
Department of Physics and Department of Materials Science and Engineering, University of Virginia, Charlottesville, VA, U.S., 22904
Prof. Jiwei Lu
Department of Materials Science and Engineering, University of Virginia, Charlottesville, VA, U.S., 22904
E-mail: jl5tk@virginia.edu

The metal insulator transition (MIT) in $VO_2$ has been an important topic for recent years. It has been generally agreed that the mechanism of the MIT in bulk $VO_2$ is considered to be a collaborative Mott-Peierls transition, however the effect of the strain on the phase transition is much more complicated. In this study the effect of the large strain on the properties of $VO_2$ films was investigated. One remarkable result is that highly strained epitaxial $VO_2$ thin films were rutile in the insulating state as well as in the metallic state. These highly strained $VO_2$ films underwent an electronic phase transition without the concomitant Peierls transition. Our results also show that a very large tensile strain along the *c*-axis of rutile $VO_2$ resulted in a phase transition temperature of ~ 433 K, much higher than in any previous report. Our findings elicit that the metal insulator transition in $VO_2$ can be driven by an electronic transition alone, rather the typical coupled electronic-structural transition.

## 1. Introduction

The metal-insulator transition (MIT) is an intriguing property of vanadium dioxide ($VO_2$), which can be of benefit in many electronic and optical applications of sensing and switching. Bulk $VO_2$ undergoes a first order phase transition from a monoclinic structure (*M1*) to a rutile structure (*R*) at ~ 340 K, which is accompanied by drastic changes in the electric and optical conductivities. The very abrupt changes in physical properties lead to many potential ultrafast optical and electrical switching applications. It has been generally agreed that the mechanism of the MIT in bulk $VO_2$ is considered to be a collaborative Mott-Peierls transition, however the effect of strain on the phase transition is much more complicated. Various types of strains such as macro-strains (mechanical strains, epitaxial strains etc.) and micro-strains via chemical substitutions in the lattice have been exploited to modify the phase transition. Recently reports revealed that single crystal $VO_2$ nanobeams under uniaxial strain exhibited a complex blend of insulating phases including the *M1*, *M2* and triclinic (*T*) phases near the MIT.[1,2] Similarly, the micro-strain in the lattice induced by the chemical doping of Al, also resulted in the meta-stable *M2* phase in free-standing $VO_2$ films near room temperature.[3] In contrast, epitaxial strain in thin film $VO_2$ from the crystal clamping with the substrate revealed a quite different picture of the phase transition. Despite the large lattice mismatch (~3.7 %) along the *c*-axis between the rutile $TiO_2$ and the $VO_2$ film, as summarized in table I, the experimental results by Muraoka *et al.*[4] showed a strong correlation between the transition temperature ($T_{MIT}$) and the uniaxial strain along the *c*-axis of $VO_2$ (*R*), and the correlation was opposite to that predicted by the Peierls (structural) mechanism.[5] Laverock *et al.*[6] have observed a Mott-like transition with a large tensile strain along the $c_R$ axis in $VO_2$ using soft x-ray spectroscopy, that demonstrated the absence of the large structural distortion near the phase transition, previously observed in bulk and moderately strained $VO_2$. Epitaxial films under bi-axial strain also demonstrated a very pronounced anisotropy in optical and transport properties,[7,8] as a result of the formation of unidirectional stripe states in which the semiconducting and metallic states coexisted.[9]

In this study, we grew epitaxial $VO_2$ films with thickness from 5~ 17 nm on various single crystal substrates to obtain very high epitaxial strains. As a result, the maximum $T_{MIT}$ reached in this study is > 430 K, which is higher than previous reports, with 4 orders of magnitude change in the resistivity. Intriguingly, Raman spectroscopy revealed that the structures of $VO_2$ films grown on $TiO_2$ in the insulating states resembled metallic rutile $VO_2$, instead of the *M1* phase observed when the films were grown on *c*-plane sapphire.

## 2. Results and Discussion

### 2.1 Microstructure characterization

In this study, epitaxial $VO_2$ films with thickness from 5~ 17 nm were grown on various single crystal substrates. The AFM images are shown in **Figure 1a**. The root-mean-square roughnesses of the films are also summarized in **Table 1**. AFM has revealed the smooth and uniform surface, and there are no cracks or pinholes. There are some signs of island growth mode of $VO_2$ for the very thin samples deposited 10 minutes or less, which is due to the large lattice mismatch between the film and the substrate. The grain size also increases with the film thickness.

The out-of-plane ($2\theta$) XRD scans showed that the (020) peaks of monoclinic $VO_2$ were coupled to the (0006) peaks of the $Al_2O_3$ substrate as seen in figure 1b, and the $VO_2$ films deposited were highly textured and $VO_2$ was the only phase detected in wide-range $2\theta$ scans, there are no other oxides grown on the substrate. As the films get thicker, the (020) $VO_2$ peaks shift to higher $2\theta$ values, which mean the lattice parameter $b$ (for monoclinic phase) gets smaller for thicker film ($b$ is reported in table 1). The $Al_2O_3$ substrate has hexagonal symmetry, and the structure of $VO_2$ deposited on top is monoclinic, which can have three preferred in-plane orientations according to the substrate and film crystal structures. The in-plane lattice spacing of bulk $VO_2$ is larger than that of $Al_2O_3$, hence an in-plane compressive strain is introduced for the film deposited on the $Al_2O_3$ substrate.

Regarding to the $VO_2$ grown on $TiO_2$ substrates, the full epitaxial relationship of the films to the substrates was obtained due to the common rutile structure for both $VO_2$ and $TiO_2$. The lattice parameters of bulk $VO_2$ are smaller that that of the $TiO_2$ substrate, hence this introduces an in-plane tensile strain due to this lattice mismatch.[10] Figure 1b also shows the out-of-plane XRD scans of the $VO_2$/(100) $TiO_2$ samples. The (200) peak in the $VO_2$ diffraction is coupled to the (200) peak of the $TiO_2$ substrate, and it is the only peak detected in the wide-range scans (not shown here). The in-plane $\phi$ scans of the (101) plane were performed to confirm the epitaxial growth of the $VO_2$.[10] In-plane $2\theta$ scans (not shown here) were also performed on the (110) and (101) planes of the substrate. The lattice parameters were directly determined from the $2\theta$ peak positions of the (200), (110), and (101) peaks of $VO_2$, though only $c$ is shown in table 1. The same processes were repeated on $VO_2$ films deposited on (001) and (011) $TiO_2$ substrates. All the lattice parameters, $a$, $b$, and $c$ are plotted as a function of thickness in the supplementary information.[10] As expected, epitaxial strains were relaxed with the increase of thickness as shown in figure 1c, and showed different relaxation rate as a function of the orientation of the $TiO_2$ substrate. The mechanism for the strain relaxation is likely due to the introduction of misfit dislocations when the thickness of the epitaxial film is larger than the critical thickness.[10]

### 2.2. Epitaxial strains and stresses

The strain and stress is analyzed using the elasticity tensor based on the rutile structure of $VO_2$, in order to understand the effect of substrate clamping on the strain/stress as a function of crystallographic orientation. For the epitaxially strained films presented here, the stress and the strain are second-rank tensors, and the Young's moduli is a fourth-rank tensor. Due to the crystal symmetry, Hooke's law can be expressed to matrix notation as shown below, with the stiffness (Young's modulus) of the tetragonal space group P$4_2$/mnm of $VO_2$ ($R$),

$$\begin{pmatrix} \sigma_1 & & \\ & \sigma_2 & \\ & & \sigma_3 \end{pmatrix} = \begin{pmatrix} E_{11} & E_{12} & E_{13} & & & \\ E_{12} & E_{11} & E_{13} & & & \\ E_{13} & E_{13} & E_{33} & & & \\ & & & E_{44} & & \\ & & & & E_{44} & \\ & & & & & E_{66} \end{pmatrix} \begin{pmatrix} \varepsilon_1 & & \\ & \varepsilon_2 & \\ & & \varepsilon_3 \end{pmatrix} \qquad (1)$$

Where $E_{ij}$ are the directional Young's moduli of $R$-$VO_2$, whereas $E_{44}$ and $E_{66}$ are for the shear stress. We transform the property tensors for analyze by choosing the coordinate system as $x_1$//[100]$_R$, $x_2$//[010]$_R$, and $x_3$//[001]$_R$ for all the samples deposited on rutile $TiO_2$. We find that $E_{11}$ ($E_{100}$) is ~ 200 GPa,[11] and $E_{12}$ ($E_{110}$) ~ 90 GPa.[12] However, there was a report on various simulated parameters, $E_{ij}$, for single crystal $VO_2$.[13] The epitaxial strains were calculated from $\frac{\Delta L_i}{L_i}$, *i.e.* $\varepsilon_1 = \frac{a - a_{bulk}}{a_{bulk}}$, where $a$ is the measured lattice parameter along $a$-axis of $VO_2$ ($R$).

For further analysis, Eqn. 1 can be reduced to:

$$\begin{aligned}
\sigma_1 &= E_{11}\varepsilon_1 + E_{12}\varepsilon_2 + E_{13}\varepsilon_3 \\
\sigma_2 &= E_{12}\varepsilon_1 + E_{11}\varepsilon_2 + E_{13}\varepsilon_3 \\
\sigma_3 &= E_{13}\varepsilon_1 + E_{13}\varepsilon_2 + E_{33}\varepsilon_3
\end{aligned} \tag{2}$$

From Hooke's law, the bi-axially strained thin film boundary condition is applied according to the substrate orientation. It is stress-free along the out-of-plane direction, i.e. perpendicular to the film surface. For $VO_2$/(100) $TiO_2$, the condition $\sigma_1 = 0$, yields $\varepsilon_3 = \frac{E_{11}(-\varepsilon_1) + E_{12}(-\varepsilon_2)}{E_{13}}$. We determined $\varepsilon_i$ from the measured lattice parameters,[10] which agrees well with this formula. Both $\varepsilon_1$ and $\varepsilon_2$ are negative (compressive strains), while $\varepsilon_3$ is positive (tensile strain). As the in-plane tensile strain $\varepsilon_3$ decreased, the out-of-plane compressive strain $\varepsilon_1$ became more negative; hence the in-plane compressive strain $\varepsilon_2$became less negative, approaching zero.

For $VO_2$/(001) $TiO_2$, the conditions $\sigma_1 = \sigma_2$, and $\sigma_3 = 0$, give $\varepsilon_1 = \varepsilon_2$, and $\varepsilon_3 = -\frac{2E_{13}}{E_{33}}\varepsilon_1$, respectively. The measured lattice parameters also agree with these conditions.[10] As the out-plane lattice parameter $c_R$ decreased, the in-plane lattice parameter $a_R$ increased. However, the measured out-of-plane strain $\varepsilon_3$ fluctuates around zero with a declining tendency, when it should be negative (compressive), as the in-plane strains $\varepsilon_1$and $\varepsilon_2$ are both positive (tensile). The result could also mean that $E_{33} >> E_{13}$.

For $VO_2$/(011) $TiO_2$, the condition $\sigma_{[011]} = 0$, results in $\sigma_2 = -\sigma_3$, hence $\varepsilon_3 = \frac{(E_{11} + E_{13})(-\varepsilon_2) - (E_{12} + E_{13})\varepsilon_1}{(E_{13} + E_{33})}$. In principle, from the substrate clamping effect, $\varepsilon_1$is positive (tensile). The measured lattice parameters showed that $\varepsilon_2$ is negative (compressive), and $\varepsilon_3$ is positive (tensile).[10] As $\varepsilon_3$ was relaxed, $\varepsilon_2$ became more compressive, then the tensile strain $\varepsilon_1$ should increase faster than the compressive strain $\varepsilon_2$ (assuming $E_{11} > E_{12}$). The trend demonstrated by the evolution in the lattice parameters as a function of the film thickness agree with these nominal conditions but the rate of change does not agree very well, which implies that $E_{12}$ may be larger than $E_{11}$.

In comparison, the stress and strain in the uniaxially strained nanobeams is very simple.[1,2] The uniaxial stress has the following stress tensor form:

$$\sigma = \begin{pmatrix} 0 & 0 & 0 \\ 0 & 0 & 0 \\ 0 & 0 & P \end{pmatrix} \tag{3}$$

Where $P$ is the mechanical stress applied along c-axis of monoclinic $VO_2$. When $P$ is tensile (positive), the strain along $c$ is tensile, and the strains along [100] and [010] are compressive. When $P$ is compressive (negative), the strains along [100], [010], and [001] change their signs accordingly.

### 2.3. Raman spectra

**Figure 2** shows Raman spectra of $VO_2/TiO_2$ compared to that of $VO_2/c\text{-}Al_2O_3$. The spectra of $VO_2/TiO_2$ shown here (before substrate/background subtraction) were dominated by the $TiO_2$ signal at room temperature and at 150 °C. The room temperature spectra of $VO_2/TiO_2$ did not reveal any insulating *M1* features, but rather showed small diffusive peaks after background subtraction, similar to what has been reported for metallic $VO_2$ (*Rutile phase*) and to what we observed in the high temperature spectra of $VO_2/c\text{-}Al_2O_3$. Even though the lattice parameter $c$ (i.e. $c$-strain) varies for the different film thicknesses, the Raman spectroscopy shows the *M1* phase only for all $c\text{-}Al_2O_3$ samples at lower temperature, and the $\omega_0$ peak shifts slightly from 615 to 617 $cm^{-1}$ as the strain along [001] *M1*-$VO_2$ was increasing. The tendency of the peak shift agrees with a previous report on $VO_2$ micro-beams,[1] but the $T$ phase and *M2* phase were not observed in any of our samples.

For the $VO_2/TiO_2$ samples, only the insulating-*R* phase was observed and there wasn't any frequency shift at room temperature. The crystal structure of the substrate, i.e. the symmetry and the lattice parameters, plays an important role in determining the symmetry of epitaxial films and the strains along different in-plane directions, hence for the $TiO_2$ substrates, only the *R*-phase $VO_2$ was observed in the given thickness range. As a comparison, the free-standing nanobeams can adopt a crystal structure either with a different symmetry (*M2*, *T*) to accommodate the applied mechanical strain.[2]

### 2.4. Metal insulator transition

**Figure 3** summarizes the transport behavior of $VO_2$ deposited on $TiO_2$ and sapphire substrates. The transport behaviors of ~13 nm $VO_2$ deposited on various substrates are shown in figure 3a. The inset shows an image of the device with Ti/Au top contacts for the measurements. The film deposited on (100) $TiO_2$ has the highest $T_{MIT}$ ~358 K, while the film on (0001) $Al_2O_3$ has a $T_{MIT}$ ~332 K, the film on (011) $TiO_2$ has a $T_{MIT}$~321 K and the film on (001) $TiO_2$ has the lowest $T_{MIT}$ ~ 305 K. A similar shift was explained using the strain from the $TiO_2$ substrate in a previous report.[4] In this current experiment, a systematic study of the thickness dependence of $T_{MIT}$ on various substrates is reported to further understand how to control the strain via different film thicknesses. The resistivities as a function of temperature for various thicknesses of $VO_2$ grown on (100) $TiO_2$ are shown in figure 3b. As the films get thinner (more strained), the $T_{MIT}$ shifts to higher temperature. The 4.9 nm film has an unusually high $T_{MIT}$ (~ 433 K) as seen in figure 3b inset. The $T_{MIT}$ were determined from resistivity as a function of temperature measurements on all other films.[10]

A phase diagram is proposed in **Figure 4** for the bi-axially strained $VO_2$. It shows the regimes of the metallic phase, the semiconductor phase, and the region of coexistence of the rutile phase, that is based on hysteresis loop and anisotropy of the transport data collected from previous reports.[7-9,14] In this phase diagram, the $T_{MIT}$ shows a rather linear relationship with the $c$-axis strain, which is in good agreement with a previous report.[4] It is worthy noting that that the lattice parameter $c$ in Muraoka's report was calculated assuming the volume of the unit cell was conserved. In this study, the lattice parameters were directly determined from XRD $2\theta$ scans. This new phase diagram extracted from the experimental data confirms the prediction, using cluster-dynamic mean field theory that investigates the effect of the epitaxial strain on the electronic structure of rutile $VO_2$,[5] where the strain dependence of $d_{||}$ state has also been confirmed by Laverock *et al.*[15] Not only does it show that the rutile phase can stabilize both metallic and insulating states, but also predicts that the coexistence of both insulating and metallic phases, for large values of the on-site Coulomb interaction potential ($U$). Experimentally, the photoemission spectroscopy on bi-axially strained $VO_2$ on $TiO_2$ substrates showed a weak insulating gap as well as the suppression of orbital redistribution across the transition, which lead to a conclusion of a more Mott-like MIT with the absence of a structural distortion, i.e. Peierls transition.[6] In highly strained epitaxial films (with the thickness much larger than what were reported here), an intriguing unidirectional strip state was observed with mixed metallic and insulating phases via scattering-type scanning near-field optical microscopy,[9] which serve as strong evidence for the coexistence of rutile insulating and metallic phases.

The films on (100) $TiO_2$ substrates have the largest strain, resulting the highest $T_{MIT}$, while the films on (001) $TiO_2$ substrates have the smallest strain, and the lowest $T_{MIT}$. Being the most strained films, the $VO_2$/(100) $TiO_2$ samples experience the largest shift in the transition temperature ($T_{MIT}$) as the thickness varies, on the other hand, the thickness has the smallest effect on the $c$-strain and the $T_{MIT}$ for the (011) and (001) samples. This is because the $c$-mismatch is much larger than the $a$-mismatch, hence the (100) samples get the largest (in-plane) strain from the substrate clamping effect, and can relax much faster than (011) and (001) samples. The $T_{MIT}$ can be modified in a wide range from 300 - 440 K, via strain manipulation by both substrate choice and thickness, which enhances the useful temperature range of the MIT for potential applications.

## 3. Conclusion

In summary, the properties of single-phase strained $VO_2$ thin films were studied on various single crystal substrates. Raman spectra showed that epitaxial $VO_2$ films grown on $TiO_2$ single crystal substrates were the rutile phase in both insulating state and metallic state, despite various orientations, instead of the *M1* state observed on *c*-plane sapphire. The strained $VO_2$ underwent an electronic phase transition without the Peierls transition as observed in single crystal and polycrystalline $VO_2$. A new phase diagram of bi-axially strained $VO_2$ was proposed, in which only the rutile $VO_2$ is presented. Secondly, a large increase in $T_{MIT}$ up to 433 K was observed for $VO_2$/(100) $TiO_2$, much higher than any previous report, while the $T_{MIT}$ of $VO_2$/(001) $TiO_2$ was reduced to 305 K, with the large, 3-4 orders of magnitude change in resistivity preserved. The ability to tune the $T_{MIT}$ using strain engineering that has been demonstrated in this work extends the potential temperature range for the use of $VO_2$ in nanoelectric devices.

## 4. Experimental Section

Reactive Target Ion Beam Deposition (RBTIBD) was used to grow $VO_2$ thin film on (100), (001), (011) $TiO_2$ and *c*-$Al_2O_3$ substrates. The systematic study of growth conditions can be found elsewhere.[16] The Raman spectroscopy with a 488 nm laser source, was conducted at room temperature and at 150 °C. To characterize the transport behavior of the films, photolithography was used to fabricate 250 × 250 $\mu m^2$ top contacts with a separation of 5 μm. The Ohmic contact was 100 nm Au / 10 nm Ti, deposited by electron beam evaporation. The temperature dependence of the *dc* resistivity was measured with a heating/cooling rate of 2 K/min, from 250 to 400 K.

**Acknowledgements**
We thank Y. Wang and L.-k. Tsui for help with the Raman spectroscopy, and J. Florro for the discussions. We acknowledge the support by the Nanoelectronics Research Initiative (NRI) and VMEC.

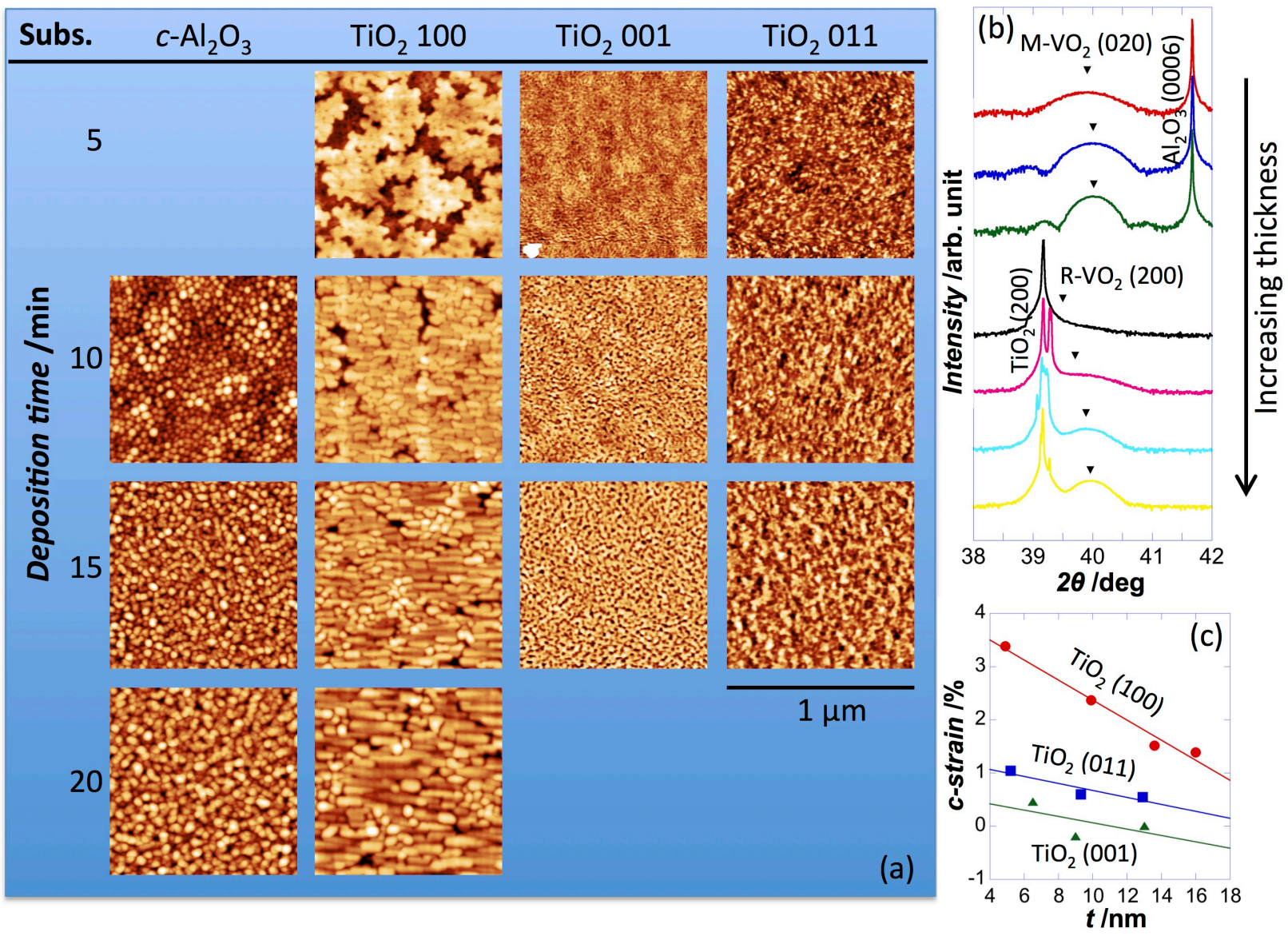

**Figure 1.** (a) 1×1 $\mu m^2$ AFM images of all samples (normalized scale for better detail). (b) $2\theta$ XRD scans of $VO_2$/*c*-$Al_2O_3$ and $VO_2$/(100) $TiO_2$, showing $VO_2$ peak couple to the substrate peak, and peak shifts as film thickness increases. (c) $c_R$ strain vs. film thickness (t) for $VO_2$/(100) $TiO_2$ samples.

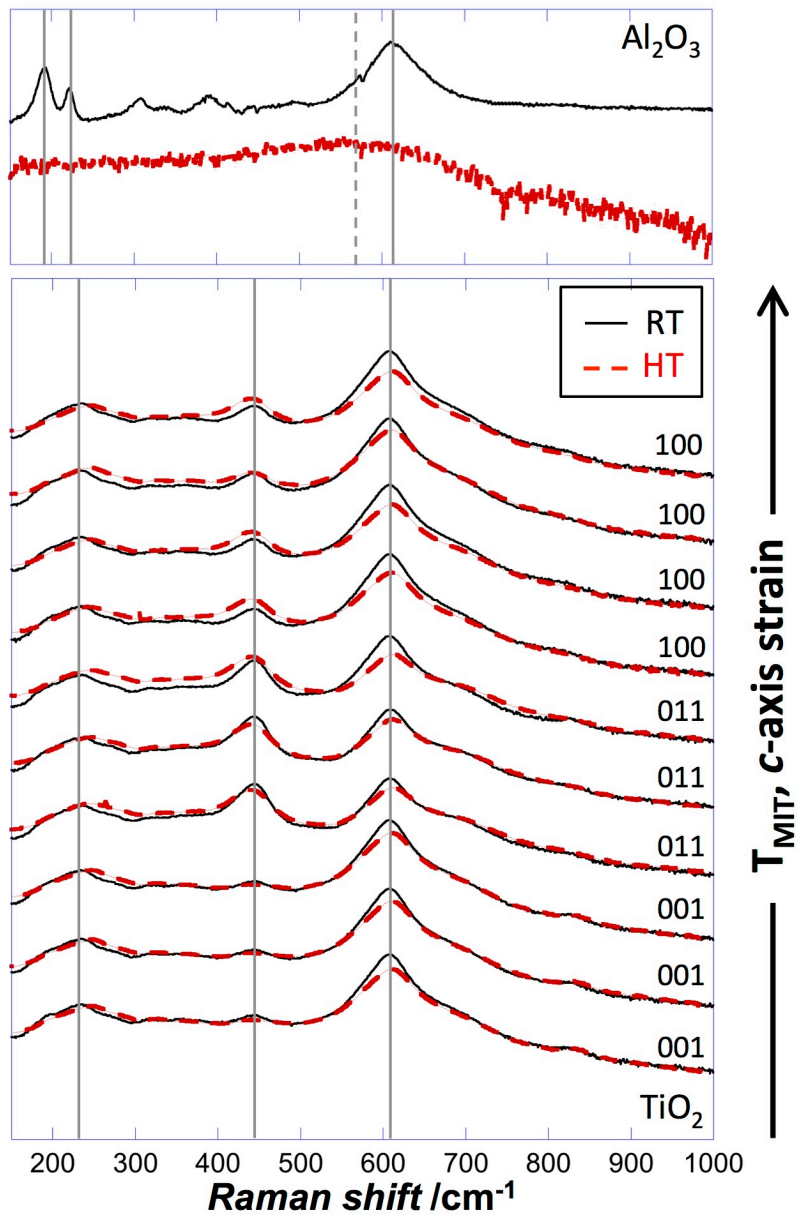

**Figure 2.** Top: Raman spectra of $VO_2$ thin film grown on *c*-$Al_2O_3$ (for reference) showing insulating *M1* structure at room temperature and showing metallic rutile structure at 150 °C respectively. Bottom: Raman spectra of $VO_2$ thin film grown on $TiO_2$, measured at room temperature and 150 °C.

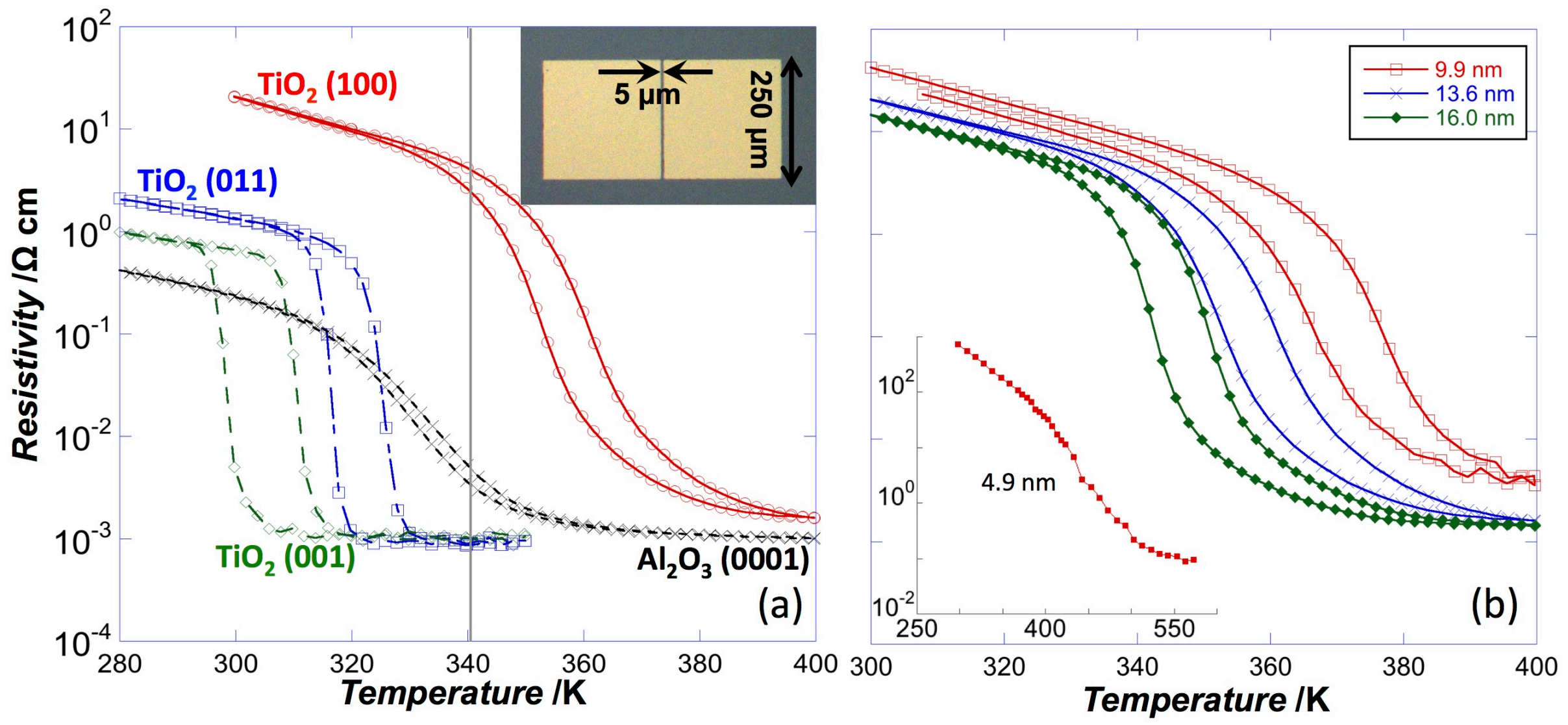

**Figure 3.** (a) Resistivity as a function of temperature of 15-min deposited (12.7-13.6 nm) $VO_2$ on various substrates, the vertical solid line indicate $T_{MIT}$ of bulk, the inset shows device image with Ti/Au top contacts and 5μm × 250μm $VO_2$ channel. (b) Resistivity as a function of temperature of various thicknesses $VO_2$/(100) $TiO_2$, the inset shows the extended temperature measurement of the thinnest film.

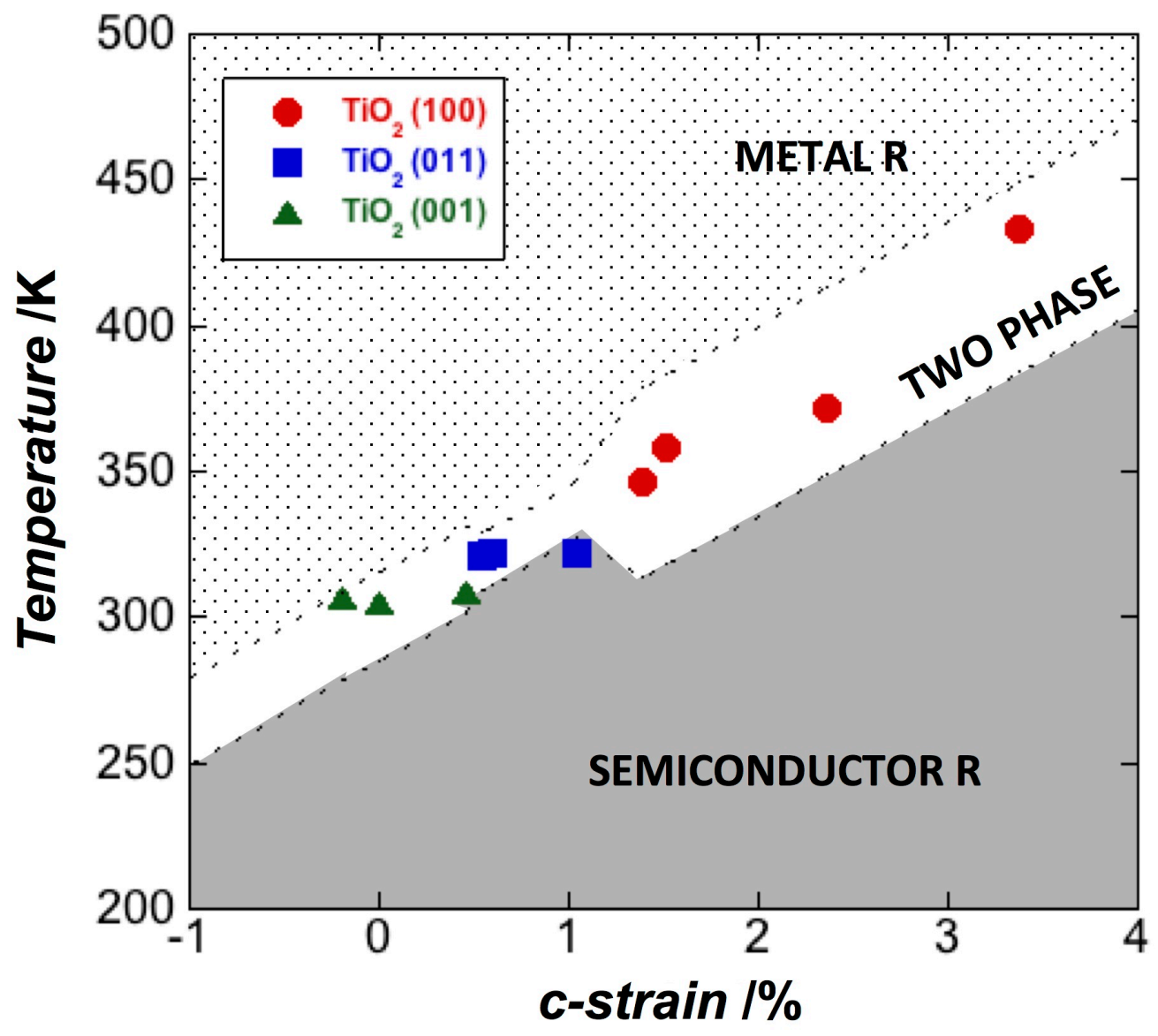

**Figure 4.** Phase diagram of rutile $VO_2$ collected from $T_{MIT}$ as a function of uniaxial *c*-strain, hysteresis loop, and transport anisotropy of $VO_2$ deposited on various $TiO_2$ substrates.

**Table 1.** Summary of thickness, RMS roughness, and lattice parameter (*b* or *c*)

| Deposit time (min) | *c*-$Al_2O_3$ | | | $TiO_2$ (100) | | | $TiO_2$ (011) | | | $TiO_2$ (001) | | |
|---|---|---|---|---|---|---|---|---|---|---|---|---|
| | *t* (nm) | RMS (nm) | *b* (Å) | *t* (nm) | RMS (nm) | *c* (Å) | *t* (nm) | RMS (nm) | *c* (Å) | *t* (nm) | RMS (nm) | *c* (Å) |
| **5** | | | | 4.9 | 0.81 | 2.9493 | 5.2 | 0.19 | 2.8826 | 6.5 | 0.12 | 2.8660 |
| **10** | 8.7 | 1.60 | 4.5132 | 9.9 | 0.88 | 2.9204 | 9.3 | 0.24 | 2.8699 | 9.0 | 0.25 | 2.8475 |
| **15** | 12.7 | 0.40 | 4.5028 | 13.6 | 1.10 | 2.8961 | 12.9 | 0.34 | 2.8684 | 13.0 | 0.49 | 2.8529 |
| **20** | 17.0 | 0.60 | 4.5014 | 16.0 | 1.45 | 2.8924 | | | | | | |

# Supporting Information

## Large epitaxial bi-axial strain induces a Mott-like phase transition in $VO_2$

*Salinporn Kittiwantanakul, Stuart A. Wolf, Jiwei Lu**

The lattice parameters of bulk $VO_2$ are smaller that that of the $TiO_2$ substrate, hence this introduces an in-plane tensile strain due to this lattice mismatch as summarized in **Table S1**. Based on the lattice mismatch, we have calculated the critical thickness ($H_c$) for $VO_2$ grown on various orientations of rutile $TiO_2$. Below the critical thickness, the epitaxial film is fully strained thanks to the mismatch. Once the thickness exceeds the critical thickness, the misfit dislocations form spontaneously to relax the epitaxial strain, hence the strain of film shows a strong dependence to the film thickness.

**Table S1.** Lattice parameters, in-plane spacing, and critical thickness of $VO_2$, $TiO_2$ (Å)

| Material | *a* | *c* | $d_{011}$ | $d_{110}$ |
|---|---|---|---|---|
| R-$VO_2$ | 4.5546 | 2.8528 | 2.4177 | 3.2206 |
| $TiO_2$ | 4.5936 | 2.9582 | 2.4871 | 3.2482 |
| Mismatch (%) | 0.86 | 3.69 | 2.87 | 0.86 |
| $H_c$ | 2.6595 | 0.3861 | 0.4211 | 1.8806 |

The out-of-plane, in-plane $2\theta$ scans and in-plane Phi ($\phi$) scans were performed to provide the lattice parameters, and confirm epitaxial growth on $TiO_2$ substrates. **Figure S1a** shows the $\phi$ scan showing $VO_2$ and $TiO_2$ peaks for the (101) plane, which confirms the epitaxial growth of $VO_2$ thin films with rutile crystal structure. The lattice parameters, *a*, *b*, and *c* were directly extracted from the three $2\theta$ scans for each of the samples grown on $TiO_2$ as shown in figure S1b-d. The lattice parameter *c* approaches the bulk value, as the film gets thicker for all films grown on $TiO_2$ substrates; while the (100) samples have the largest uniaxial strain along <001>, the strain decreases for the (011) samples, and the (001) samples have the least strain.

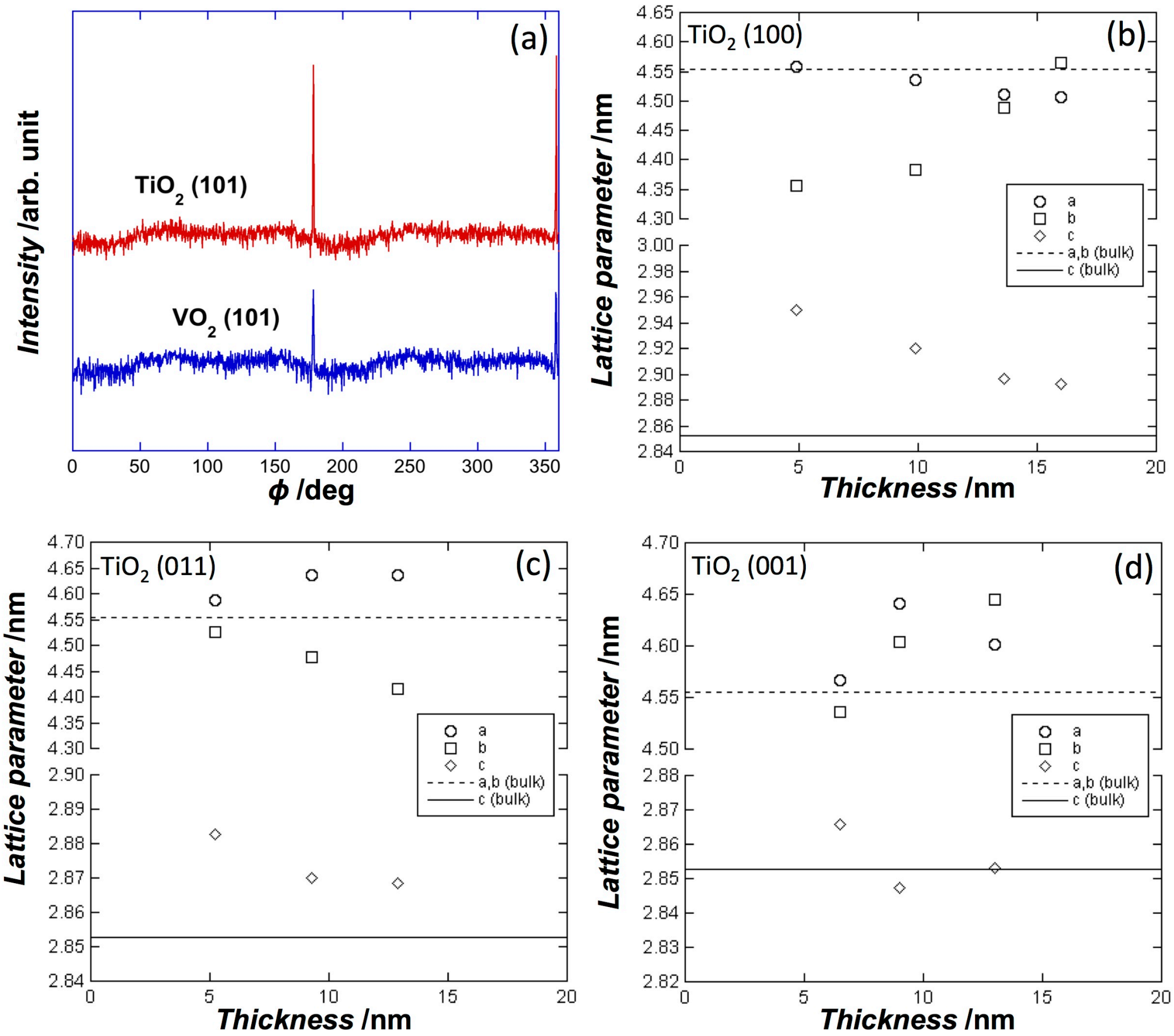

**Figure S1.** (a) The in-plane $\phi$ scans of (101) $VO_2$ and (101) $TiO_2$. (b)-(d) Lattice parameters as a function of film thickness for $VO_2$ deposited on (100), (011), and (001) $TiO_2$ respectively.

Raman spectroscopy revealed that the structure of insulating $VO_2$ films grown on $TiO_2$ resembled metallic rutile $VO_2$, instead of the *M1* structure seen on *c*-plane sapphire. The spectra of $VO_2/TiO_2$ shown here (before substrate/background subtraction) were dominated by the $TiO_2$ signal. Even though the *c* lattice parameter (i.e. *c*-strain) varies as a function of film thickness for all of the films grown on *c*-$Al_2O_3$, the Raman spectroscopy shows only the *M1* phase, and the $\omega_o$ peak shifts slightly from 615 to 617 $cm^{-1}$ as the *c*-strain increases, the peaks are summarized in table II. The tendency of the peak to shift agrees with a previous report on $VO_2$ microbeams [Atkin, J. M. *et al. Phys. Rev. B* **85**, 020101 (2012).], however the *T* phase and *M2* phase were not observed in our thin films. For the $VO_2/TiO_2$ samples, only the insulating-*R* phase was observed and does not reveal any shift at room temperature as summarized in **Table S2**.

**Table S2.** Raman shifts ($cm^{-1}$) of $VO_2$ samples

| Substrate | Thickness [nm] | Raman shifts [$cm^{-1}$] | | | | *c* or *b* [Å]* | % "*c*-strain" |
|---|---|---|---|---|---|---|---|
| $TiO_2$ 100 | 4.9 | 142 | 232 | 446 | 611 | 2.9493 | 3.38 |
| | 9.9 | 142 | 232 | 446 | 611 | 2.9204 | 2.37 |
| | 13.6 | 142 | 232 | 445 | 611 | 2.8961 | 1.52 |
| | 16.0 | 142 | 233 | 446 | 611 | 2.8924 | 1.39 |
| $TiO_2$ 011 | 5.2 | 142 | 232 | 446 | 611 | 2.8826 | 1.05 |
| | 9.3 | 142 | 233 | 446 | 612 | 2.8699 | 0.60 |
| | 12.9 | 142 | 232 | 446 | 612 | 2.8684 | 0.55 |
| $TiO_2$ 001 | 6.5 | 142 | 231 | 447 | 611 | 2.8660 | 0.46 |
| | 9.0 | 142 | 233 | 445 | 611 | 2.8475 | -0.19 |
| | 13.0 | 142 | 233 | 446 | 610 | 2.8529 | 0.004 |
| $Al_2O_3$ | 8.7 | 195 | 224 | ***615*** | 750 | 4.5132 | 0.17 |
| | 12.7 | 192 | 224 | ***616*** | 750 | 4.5028 | 0.63 |
| | 17.0 | 193 | 224 | ***617*** | 749 | 4.5014 | 0.69 |

* For samples grown on sapphire, the values are for out-of-plane 'b' for *M1* phase. For others, the values are for "c" lattice constant of *R* phase.

The temperature dependence of the *dc* resistivity was measured using a Versa-lab system (Quantum Design), with a heating/cooling rate of 2 K/min, from 280 to 400 K for $VO_2$/*c*-$Al_2O_3$, 300 to 400 K for $VO_2$/(100) $TiO_2$, and 250 to 400 K for $VO_2$ deposited on (001) and (011) $TiO_2$ substrates. The *dc* resistivity was then calculated according to the device geometry and the thickness of the film, and plotted as a function of temperature as shown in **Figure S2**. The Metal Insulator Transition Temperature ($T_{MIT}$) of each sample was extracted from the derivative of the logarithm of the resistivity, that is defined as

$$T_{MIT} = \frac{T_{up} + T_{down}}{2}$$

when $T_{up} = T$ where $\frac{d(\log \rho_{up})}{dT}$ is at a minimum, and $T_{down} = T$ where $\frac{d(\log \rho_{down})}{dT}$ is at a minimum. $\rho_{up}$ is the resistivity from the up-sweep (increasing temperature from 300 to 400 K), and $\rho_{down}$ *is* the resistivity from the down-sweep (decreasing temperature from 400 to 300 K).

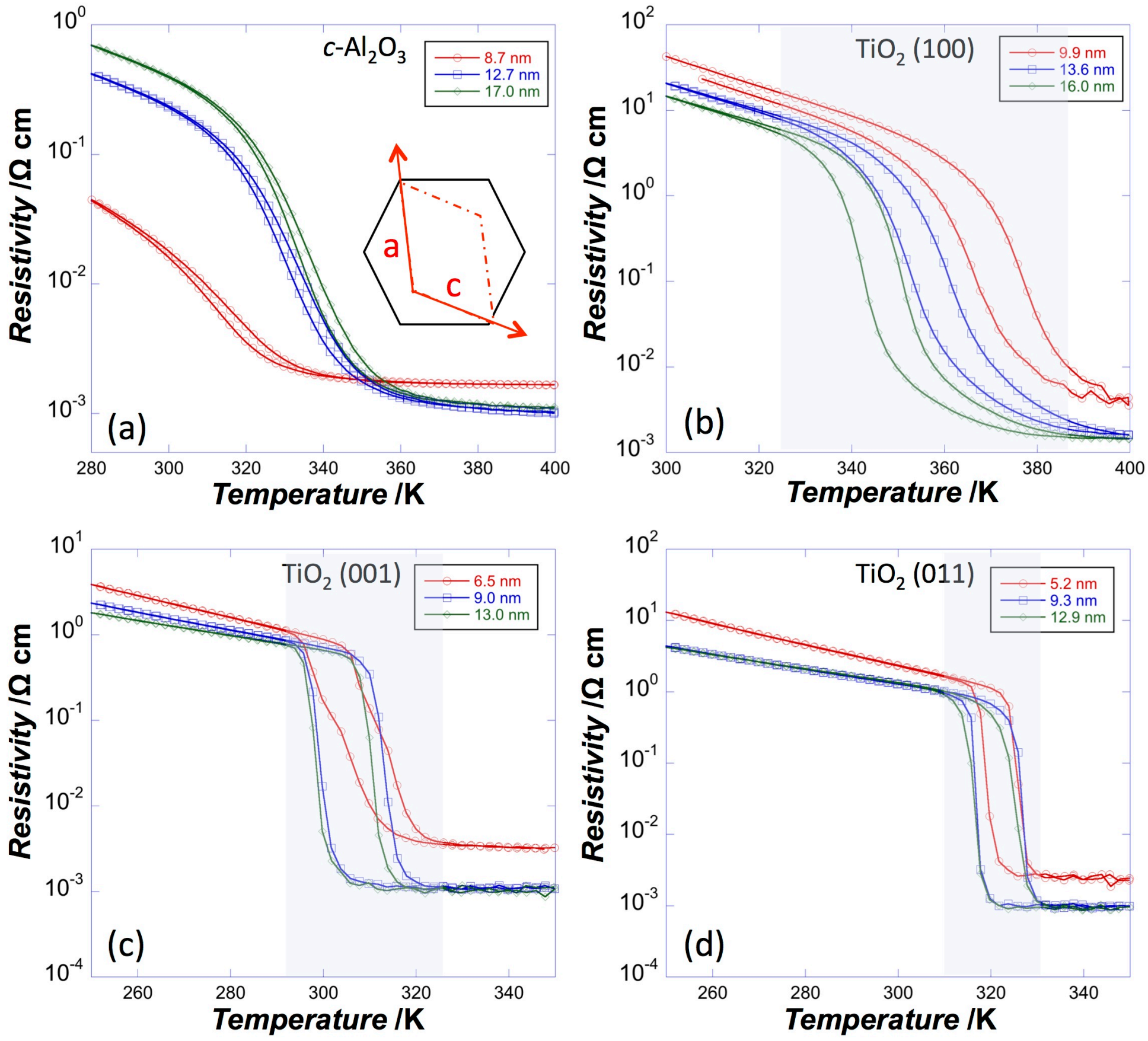

**Figure S2.** (a)-(d) Resistivity as a function of temperature of various thickness $VO_2$ deposited on *c*-$Al_2O_3$, (100), (001), and (011) $TiO_2$ respectively, inset showing monoclinic structure of $VO_2$ (red dash line) on top of hexagonal crystal structure of *c*-$Al_2O_3$ (black solid line), the highlighted area showing a possible two phase region.

All the $TiO_2$ samples preserve the resistivity ratio during the transition very well, ~3-4 orders of magnitude, while the *c*-$Al_2O_3$ samples experience great suppression as the film get thinner as shown in figure S2. This due to the fact that the films grown on *c*-$Al_2O_3$ substrates are not epitaxial, since the crystal structure of the substrate is hexagonal, and the crystal structure of $VO_2$ is monoclinic (as shown in figure S2a inset). The different crystal structures, together with the large lattice mismatch, results in poorer $VO_2$ crystallinity as compared to the epitaxial $VO_2$ grown on $TiO_2$ substrates.

Being the most strained films, the $VO_2$/(100) $TiO_2$ samples experience the largest shift in the transition temperature ($T_{MIT}$) as the thickness varies, on the other hand, the thickness has a minimal effect on the *c*-strain and the $T_{MIT}$ for (011) and (001) samples. For the (100) samples the *c*-mismatch is much larger than the *a*-mismatch, hence these samples get the largest (in-plane) strain from the substrate clamping effect, and thus relax much faster than the (011) samples and the (001) samples. Being less strained, thus better crystallinity, the (011) samples and the (001) samples show much sharper transitions then the (100) or the *c*-$Al_2O_3$ samples. There is also a signature of more defects in the thinner films, as the resistivity in the metallic phase is higher than the thicker films as seen in figure S2. Interestingly, the 6.5 nm (001) film also shows a two-step transition, suggesting multiple phases with different transition temperatures, while this is not the case for the (011) film deposited at the same condition.

The phase diagram presented in the manuscript was constructed according to previous reports on transport anisotropy and the highlighted area of coexistence as shown in Figure S2; i.e. the coexistence temperature range of the (100) $TiO_2$ samples is about ±30 K, note that the resistivity measurement was done along the *c*-axis in which the $T_{MIT}$ is 3-5 K higher than the $T_{MIT}$ along the in-plane *a*-axis, resulting in a 30 K lower bound and 35 K upper bound for the two phase region.